\documentclass[%
twocolumn,
showpacs,preprintnumbers,
 amsmath,amssymb,
 aps,
]{revtex4-1}

\usepackage{graphicx}
\usepackage{dcolumn}
\usepackage{bm}
\usepackage{upgreek}

\begin{document}

\preprint{APS/123-QED}

\title{Electron transport across a metal/MoS$_2$ interface: dependence on contact area and binding distance}

\author{Zhaoqiang Bai$^{1,2}$}%
\author{Troels Markussen$^2$}%
\author{Kristian S. Thygesen$^{2,3}$}
\email{thygesen@fysik.dtu.dk}%

\affiliation {$^1$Graphene Research Centre, Department of Physics, National University of Singapore, Singapore 117542, Singapore}%
\affiliation {$^2$Center for Atomic-scale Materials Design (CAMD), Department of Physics, Technical University of Denmark, DK - 2800 Kongens Lyngby, Denmark}%
\affiliation {$^3$Center for Nanostructured Graphene (CNG), Technical University of Denmark, DK - 2800 Kongens Lyngby, Denmark}%
\date{\today}%

\begin{abstract}
We investigate the nature of electron transport through monolayer molybdenum dichalcogenides (MoX$_2$, X=S, Se) suspended between Au and Ti metallic contacts. The monolayer is placed ontop of the close-packed surfaces of the metal electrodes and we focus on the role of the metal-MoX$_2$ binding distance and the contact area. Based on \emph{ab initio} transport calculations we identify two different scattering mechanisms which depend differently on the metal-MoX$_2$ binding distance: (i) An interface resistance between the metal and the supported part of MoX$_2$ which decreases with decreasing binding distance and increasing contact area. (ii) An edge resistance across the 1D interface between metal-supported and free-standing MoX$_2$ which increases with decreasing binding distance and is independent on contact area. The origin of the edge resistance is a metal-induced potential shift within the MoX$_2$ layer. The optimal metal thus depends on the junction geometry. In the case of MoS$_2$, we find that for short contacts, L$<$6 nm, Ti electrodes (with short binding distance) gives the lowest resistance, while for longer contacts, Au (large binding distance) is a better electrode metal.

\end{abstract}

\pacs{73.22.-f, 73.63.-b}%
\maketitle

Atomically thin, two-dimensional (2D) materials are presently being intensely researched due to their unique and easily tunable electronic properties which make them interesting candidates for the next generation nano-electronic devices. The most widely studied 2D material is graphene,\cite{01,02} and its rich physics\cite{03,04,05} and extraordinary properties such as high mobility\cite{06} and thermal conductivity are well understood by now. However, pristine graphene does not have a band gap, a property that is essential for many applications, including field effect transistors (FETs). It is possible to open small band gaps in graphene, e.g., by means of nanostructuring\cite{07,08} or substrate interactions\cite{09}, however, this inevitably leads to increased fabrication complexity and either reduces mobility to the level of strained silicon films\cite{10,11,12,13,14,15} or requires high voltages\cite{16,17}.

The \textquotedblleft band gap\textquotedblright ~problem can be overcome by using transition metal dichalcogenides, of which the molybdenum dichalcogenides (MoX$_2$, X=S, Se and Te) are a subset.\cite{18,19} In the bulk, the transition metal dichalcogenides consist of covalently bonded hexagonal layers held together by weak van der Waals forces. This allows single layers to be isolated by exfoliation techniques similar to those used to produce graphene. Monolayer MoS$_2$ has a direct (quasiparticle) band gap of around 2.5 eV\cite{20} (the optical gap is around 0.5 lower due to strong excitonic effects) and a maximum reported mobility of 517 cm$^{2}$~/(V s).\cite{21,22} Single and multilayer MoS$_2$ FETs with on/off-current ratios as high as 10$^{8}$ and steep subthreshold swing (74 mV/decade) have already been demonstrated in a top-gated transistor architecture.\cite{23} Moreover, initial studies indicate that MoS$_2$ may also be useful for sensing and energy-harvesting applications.\cite{24,25} One of the major performance-limiting factors for any nano-scale device, including atomically thin 2D materials, is interfaces where losses and back scattering inevitably takes place. In the context of FET applications, previous work has focused on the interface between the gate dielectric and the 2D semiconductor channel,\cite{26,27} while less attention has been paid to the metal/semiconductor contacts at the source and drain ends.\cite{28,29} However, high contact transparency reduces the required bias voltages for operation and thus the problem should be carefully considered when designing 2D electronic devices. Popov \emph{et al}. proposed that Ti might be more suitable as electrode material for monolayer MoS$_2$ than the commonly used Au due to the comparatively higher electron injection efficiency.\cite{30} They suggested that the nature of the contact, whether tunneling or Ohmic, plays a crucial role for the device performance. However, these conclusions were based on analysis of the electrostatic potential and density of states at the metal/MoS$_2$ interface, and no explicit calculations of the electron transport across the interface were performed.

In this paper we present a detailed investigation of the charge transport properties of MoX$_2$ monolayers suspended between different types of metal electrodes using \emph{ab initio} transport calculations.  Our calculations show that only when the overlap region between the 2D sheet and the electrode surface is less than a few nanometers, does a stronger metal-MoX$_2$ coupling lead to lower contact resistance. For larger, and practically more relevant, contact areas, a weaker coupling provides a more transparent, i.e. less resistive, interface. Specifically, we find that Au electrodes generally produce lower contact resistance compared to Ti electrodes, in direct contrast to previous predictions based on an analysis of the density of states in the interfacial region. This surprising effect is due to a larger mismatch of the electron potential in the supported and freestanding MoX$_2$ segments, respectively, when the coupling is stronger.

All electron transport calculations were performed using density functional theory and non-equilibrium Green\textquoteright s functions (DFT-NEGF) as implemented in the Atomistix ToolKit package (ATK).\cite{31,32} For the analysis of binding energies, charge transfer and effective potentials, we used the electronic structure code GPAW.\cite{33} The Perdew-Burke-Ernzerhof (PBE) exchange-correlation functional\cite{34,35} was employed throughout this work. For both ATK and GPAW calculations, the electron wave functions were expanded in a single-$\zeta$ polarized basis set. Convergence tests showed that the more accurate double-$\zeta$ polarized basis set yields essentially the same transmission functions. A kinetic energy cutoff of 150 Ry was used for representing the density and potentials, and a $1 \times 100$ Monkhorst-Pack k-point mesh\cite{36} was used to sample the transverse Brillouin zone in the transport calculations.

\begin{figure}
\centering
\includegraphics[width=0.5\textwidth]{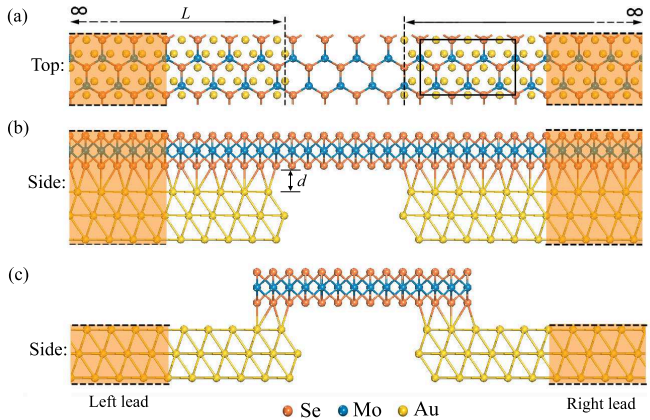}
\caption{Schematic device model of the Au/MoX$_2$ junctions. Panels (a) and (b) show the top and side view of the junction with infinite contact length L, respectively. The interlayer distance between the Au electrodes and the MoX$_2$ sheet, \emph{d}, is marked in panel (b). Panel (c) shows the junction geometry with finite contact length.}
\end{figure}

The structure of the investigated metal/MoX$_2$ junctions is shown in Fig.1. A central region is connected to two semi-infinite leads displayed by the shaded areas. Two generic types of geometries are used to model the interface corresponding to an infinite (Figs. 1(a) and 1(b)) and finite contact region (Fig. 1(c)) between monolayer MoX$_2$ and the metal surface, respectively. For the infinite contact, the MoX$_2$ sheet is included as part of the electrode. We note in passing that previous computational work on carbon nanotubes and graphene connected to metal electrodes found that the infinite contact setup generally leads to higher transparency compared to the finite contact setup, consistent with the results reported here.\cite{37,38,39}  The Au surface is modeled by a three-layer (111) slab. We have verified that the transmission function is essentially unchanged when the slab thickness is increased to six layers. Figure 1(a) shows a top view of the Au/MoSe$_2$ (111) junction with the unit cell encircled by the black solid square: Two of the bottom Se atoms are aligned on top of the surface Au atoms with the other four placed in the hollow sites, which leads to a negligible lattice mismatch (approximately 0.9\% strain applied on Au). To test the dependence of contact resistance on the metal-MoSe$_2$ coupling, we have considered different junction structures with varying binding distance \emph{d} (2, 2.25, 2.5, 2.75, and 3 {\AA}). The variations may to a first approximation mimic the binding to different kinds of metals with different metal-MoX$_2$ coupling strength. In addition to this we note that various approximations to the exchange-correlation energy within DFT give quite different equilibrium binding distances. In particular the local density approximation (LDA) is found to give a binding distance of 2.7 {\AA}, while PBE predicts 3.0 {\AA}, and a van-der-Waals functional\cite{40} yields 3.2 {\AA}. We are not aware of any experimental data or higher level calculations reported for the Au-MoSe$_2$ binding distance, and we therefore prefer to treat the distance as an adjustable parameter.

\begin{figure}
\centering
\includegraphics[width=0.5\textwidth]{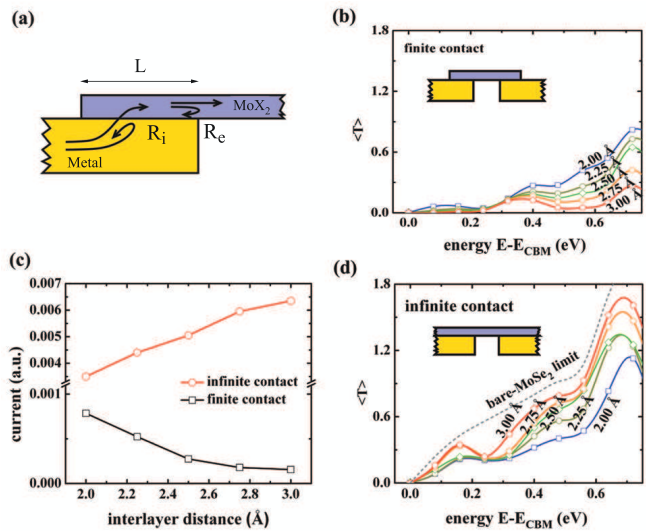}
\caption{(a) Schematic illustration of the two sources of scattering resistance. The interface resistance, \emph{R$_i$}, between the metal and MoX$_2$, is reduced by increasing the electronic coupling and the contact length L. The back-scattering taking place at the edge between supported and freestanding MoX$_2$ gives rise to an edge resistance, \emph{R$_e$}, which, on the contrary, increases with increasing electronic coupling. Panels (b) and (d) show the interlayer-distance dependence of the transmission spectra of the Au/MoSe$_2$ junctions with finite and infinite contact geometries, respectively. Panel (c) shows the current per transverse line segment at the bias of 0.1 volt with respect to the Au-/MoSe$_2$ interlayer distance for the junctions with finite (black-square line) and infinite (red-circle line) contact lengths, respectively.}
\end{figure}

Before presenting the results we consider the schematic interface shown in Fig. 2 (a). In this transport setup, an electron impinging on the junction from the left electrode have to cross two interfaces on its way to the freestanding MoX$_2$ sheet: First, the electron can back-scatter at the MoX$_2$-metal interface giving rise to an interface resistance \emph{R$_i$}. It is reasonable to expect \emph{R$_i$} to be smaller in systems having a large interlayer electronic coupling which is equivalent to a smaller metal-MoX$_2$ binding distance. Next, the electrons within the MoX$_2$ channel may become back-scattered at the metal edge when propagating from the supported to the freestanding part of the MoX$_2$ sheet, leading to an edge resistance, \emph{R$_e$}. This edge resistance originates from the change in the effective potential at the border between the supported and freestanding sheet. We show below that the edge resistance, as opposed to \emph{R$_i$}, increases with increasing electronic coupling (i.e. decreasing binding distance).

We now return to our prototypical model and the \emph{ab initio} calculations. Figure 2 (b) and 2 (d) present the evolution of the transmission spectra with respect to the interlayer distances of the Au/MoSe$_2$ junctions with finite and infinite contact area, respectively. The transmission function is defined as $\langle T(E) \rangle=\mathrm{Tr}[G^r(E)\mathit{\Gamma}_\mathrm{L}(E)G^a(E)\mathit{\Gamma}_\mathrm{R}(E)]$, where $G^{r(a)}(E)$ is the retarded (advanced) Green’s function, and $\mathit{\Gamma}_\mathrm{{(L,R)}}(E)=i(\mathit{\Sigma}_\mathrm{{(L,R)}}^r(E)-\mathit{\Sigma}_\mathrm{{(L,R)}}^a(E))$ describes the level broadening due to coupling to the left and right electrodes expressed in terms of the electrodes self-energies $\mathit{\Sigma}_\mathrm{{(L,R)}}(E)$. The distance between the two electrodes corresponds to four MoSe$_2$ unit cells, and we have checked that the transmission is converged with respect to this distance. We focus on the transmission for energies near the conduction band minimum (CBM) of MoSe$_2$ based on the experimentally observed n-type FET characteristics of the MoS$_2$ transistors, which indicates that the actual Fermi levels (E$\mathrm{_F}$) of a series of metals (including Au and Ti) line up close to the conduction band edge of MoS$_2$.\cite{28} In fact our calculations also predict that E$\mathrm{_F}$ is situated close to the MoSe$_2$ and MoS$_2$ CBMs, in agreement with previous calculations.\cite{30} For clarity, the transmission curves are shifted to align the MoSe$_2$ CBM for the different binding distances (these may vary due to the different interface dipoles formed). For a finite contact area, the transmission increases as the binding distance is decreased in agreement with intuitive expectations. In contrast, for the infinite contact geometry, the transmission increases with increasing binding distance. As we show below these trends can be explained by the relative size/importance of the interface resistance, \emph{R$_i$}, and edge resistances, \emph{R$_e$}, in the two cases.

We further evaluate the \emph{I-V} characteristics by the Landauer-Buttiker formula
\begin{equation}
I  = \mathrm{\frac{2e}{h}} \int_{-\infty}^{+\infty} \,\mathrm{d}E \langle T(E) \rangle [f(E-\mu_\mathrm{L})-f(E-\mu_\mathrm{R})],
\end{equation}
where $f(E)=1/[\mathrm{exp}⁡(E/(k_\mathrm{B}T))+1]$ is the Fermi-Dirac distribution function and $\mu_\mathrm{{(L/R)}}$ the chemical potential in the left/right electrode. The bias voltage is defined as $V=(\mu_\mathrm{L}-\mu_\mathrm{R})/e$. Figure 2 (c) illustrates the current per transverse line segment at the bias of 0.1 volt as a function of the interlayer separation \emph{d} for the junctions with finite (black square) and infinite (red circle) contacts. Again, the currents of the two geometries show an opposite dependence on the binding distance: A monotonously decreasing current is found for the finite contact geometry due the weakened Au-MoSe$_2$ bond which prevents the electron injection, whereas the opposite trend is observed for the infinite contact geometry.

The above results indicate that for the finite contact geometry, the transport is governed by the interface resistance \emph{R$_i$} which is reduced when the binding distance is decreased. We may assume that the interface resistance can be written as $R_i=\rho_i/L$, where $\rho_i$ is an interfacial resistivity and \emph{L} the length of the contact region in the transport direction. For sufficiently long contact regions we thus expect \emph{R$_i$} to be negligible and the edge resistance \emph{R$_e$} remains to be the only determining factor for the electron transmission. This explains the somewhat counterintuitive reversal of transmission in the infinite contact case: when the MoSe$_2$ layer is kept apart from the metal leads by an adequate distance, the states within the supported part of MoSe$_2$ become very similar to those within the freestanding region due to the much weakened perturbation induced by the metal electrodes. Consequently, the intrinsic edge resistance at the Au edges is eliminated, which gives rise to an enhanced transmission.
As concrete examples of strongly and weakly coupled junctions, we consider the cases of Ti/MoS$_2$ and Au/MoS$_2$, respectively. The binding distances are set to \emph{d}=2.0 and \emph{d}=2.62 {\AA}, respectively, based on previously published PBE values.\cite{30} In Ref. 30 Ti was proposed as a more promising electrode metal than Au due to its stronger bond strength and higher local density of states at E$\mathrm{_F}$, suggesting a low-resistance Ohmic contact with efficient electron injection. These predictions are indeed in agreement with our calculations for a short contact area. Figure 3 (a) shows the transmission function of MoS$_2$ connected via a short contact to a single Au and Ti electrode, respectively. We observe in general larger transmission for Ti leads reflecting the larger electronic coupling. However, the infinite contact geometry, which might represent a more realistic model of an experimental setup with contact lengths as long as hundreds of nanometers,\cite{23,28,41} exhibits the opposite tend, as shown in Fig. 3 (b). In this case, the weakly coupled Au/MoS$_2$ junction shows enhanced transmission compared with that of Ti/MoS$_2$. This is consistent with the picture obtained by varying the interlayer coupling strength in two-terminal Au/MoSe$_2$.

\begin{figure}
\centering
\includegraphics[width=0.5\textwidth]{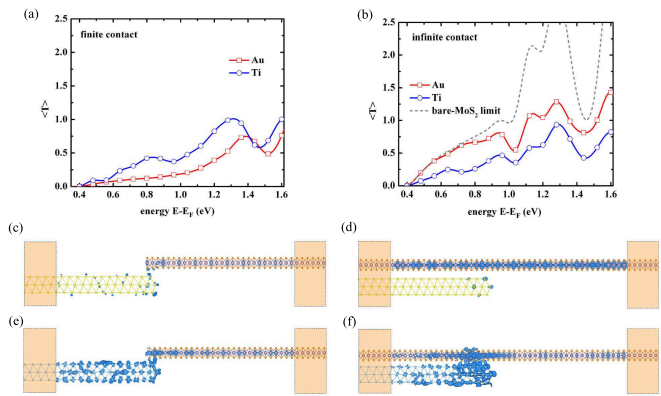}
\caption{Transmission spectra of the Au/MoS$_2$ and Ti/MoS$_2$ junctions with (a) finite and (b) infinite contact areas. Panels (c) and (d) show the transmission eigenchannels of the Au/MoS$_2$ junctions with finite and infinite contact areas, respectively. Panels (e) and (f) show the transmission eigenchannels of the Ti/MoS$_2$ junctions with finite and infinite contact areas, respectively.}
\end{figure}

To illustrate the effect of the two scattering mechanisms behind \emph{R$_i$} and \emph{R$_e$}, we plot in Figs. 3 (d) and 3 (f) the scattering states (or transmission eigenchannels)\cite{42} at the Γ point and the energy of 0.16 eV above the CBM for both the Au and Ti electrodes with infinite overlap. Again we stress that for clarity we use here a one- rather than two-terminal setup in order to eliminate the effect of interference between states reflected at the two electrodes. It is found that the eigenchannels in both cases are mainly composed of the Mo $d_{z^2}$ states. In the case of Ti electrodes, the scattering state has a large component localized at the edge of the Ti slab, which we attribute to the significant back scattering taking place within the MoS$_2$ and causing an enhanced edge resistance. In the case of Au, the scattering state is less perturbed and remains more evenly distributed throughout the junction. This is also reflected by the higher transmission value of 0.38 compared with the value of 0.15 for Ti. On the contrary, the transmission eigenchannels associated with the short contact geometry, as shown in Figs. 4 (c) and 4 (e), reveal the opposite trend. In this case, the Ti electrode yields higher transmission due to the stronger binding which facilitates a more efficient electron injection into the MoS$_2$ channel. These observations confirm that the interface resistance controls the properties of the short contact geometry, whereas the edge resistance dominates for the longer contact areas.

\begin{figure}
\centering
\includegraphics[width=0.5\textwidth]{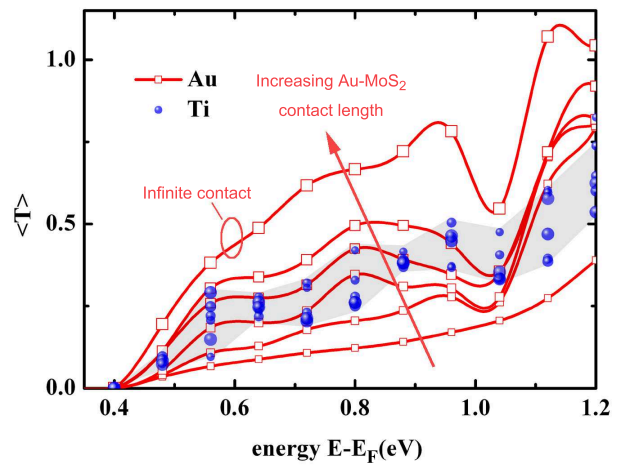}
\caption{Contact-length dependence of transmission of single-contact Au/MoS$_2$ (red square) and Ti/MoS$_2$ (blue circle) junctions. Different sizes of symbols denote different contact lengths of 0.25, 0.5, 2.5, 4.5, and 6.5 unit cells, as well as the infinite contact, with larger symbols representing longer contacts. The shaded area marks a rough range where the transmission of the Ti/MoS$_2$ junctions may vary.}
\end{figure}

To investigate the influence of the contact region on the transport properties we show in Fig. 4 the transmission function for energies corresponding to the bottom of the conduction band for various sizes of the contact region. For the Au electrode the transmission clearly increases with increasing metal/MoS$_2$ overlap. In contrast, for the Ti electrode, the transmission is essentially independent of the size of the overlap. Again, this behavior is consistent with the weakly (strongly) coupled junction being dominated by the interface (edge) resistance. Specifically, for the weakly bonded Au/MoS$_2$ junction, transport across the 2D interface is the bottleneck. As mentioned above, \emph{R$_i$}  becomes negligible when the contact is sufficiently long, and therefore, the transmission monotonically approaches the limit of the bare MoS$_2$ value as the contact region increases. For the strongly bonded Ti/MoS$_2$ junction, however, the edge resistance, which is much less sensitive to the contact length, dominates the transport. This explains the rather invariant transmission of Ti/MoS$_2$ with respect to the overlap size. From the results in Fig. 4 we conclude that the question of \textquoteleft which metal has the smallest contact resistance?\textquoteright ~only can be answered rigorously, if the contact length is specified. Upon inspection of Fig. 4 we find that for contact lengths above 6.5 unit cells (
$\scriptsize{\sim}$ 6.2 nm) the transmission through the Au junction exceeds that of Ti in most of the energy range near the CBM. From this we estimate that for short overlap lengths (L$<$6.2 nm) Ti provides a more transparent contact to MoS$_2$ while for longer contact lengths Au is a superior electrode material.

\begin{figure}
\centering
\includegraphics[width=0.5\textwidth]{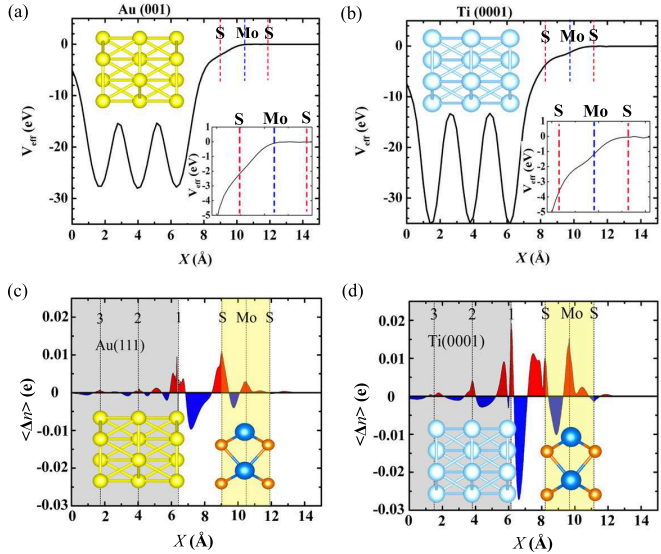}
\caption{Effective potential for bare Au (a) and Ti (b) slabs along the surface normal, averaged over the transverse directions. The dashed lines mark the position of the S and Mo layers. Note that the first S layer experiences a significant potential. The same is true for the Mo layer at Ti substrate. Panels (c) and (d) show the electron density difference for Au (c) and Ti (d) substrates averaged over the two transverse directions. The induced charges partly screen the metal potential on the Mo- and first S-layers.}
\end{figure}

In order to shed light on the origin of the edge resistance, we consider in Fig. 5 (a) and (b) the effective potential from the pure Au and Ti slabs, respectively. The potentials are averaged over the transverse directions. The potentials have minima at the planes of the metal ions, and decays in the vacuum region. With dashed lines we show the positions of the Mo and two S layers, when the interface is formed. From the insets, which show a zoom in around the MoS$_2$ layers, it is clear, that electrons in the closest (lower) S-layer feel a significant attractive potential from the metal ions lowering their potential energy with 2-4 eV. On the other hand, the potential is essentially zero at a distance corresponding to the upper S-layer, and we expect no direct effect of the metal potential. In the case of Ti substrate, electrons in the Mo layer also experience an attractive potential of 1-2 eV. These potential values correspond to the unscreened, bare metal potential.
When the metal/MoS$_2$ interface is formed, charge will be redistributed forming surface dipoles. In Fig. 5 (c) and (d) we show the charge density difference between infinite MoS$_2$ on Au (c) and on Ti (d) relative to the isolated MoS$_2$ and metal systems. In agreement with previous works\cite{30,43} net electron transfer is found from the metal to the MoS$_2$ layer, with comparatively larger amount depleted from Ti than that from Au. The redistributed charge leads to screened potential, where the effects of the metal potential in the MoS$_2$ layers are reduced: The net charge accumulation around the first S-layer increases the electron potential, thus working in the opposite direction than the bare (unscreened) metal potential. Based on these results one may expect that the potential within the MoS$_2$ layer varies significantly between metal supported and unsupported region.

\begin{figure}
\centering
\includegraphics[width=0.5\textwidth]{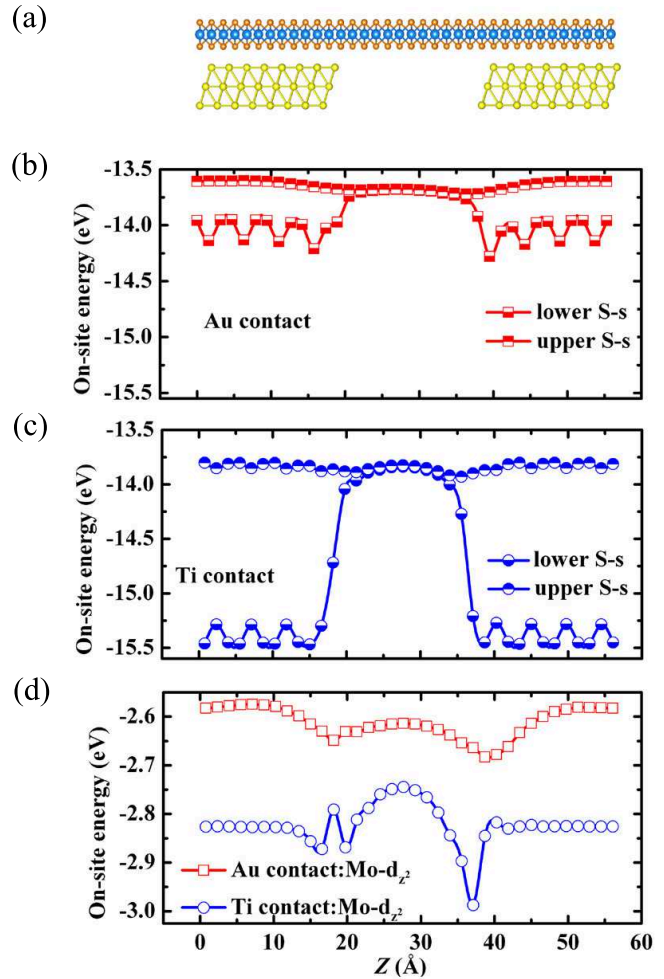}
\caption{Potential variations within MoS$_2$ along the transport. Panels (b) - (d) show variations of orbital (on-site) energies along the transport direction indicated in panel (a). The s-orbital energies of the upper and lower sulfur layers are shown in panels (b) and (c) for Au and Ti substrates, respectively. The variations of the, primary current carrying, Mo $d_{z^2}$ orbitals is shown in (d) for both Ti and Au substrates.}
\end{figure}

In Fig. 6 we analyze the potential variation along the transport direction for the junction structure shown in panel (a). These potential variations are the physical origin of the edge resistance. Panels (b) and (c) show the potential, as probed by the sulfur s-orbital energies, in the upper and lower sulfur layers is shown along the transport direction with Au and Ti leads, respectively. For the lower sulfur layer we observe a potential hill in the gap region reflecting that the influence by the metal ionic potentials is the dominant effect close to the metal: The metal ionic potential lowers the energy in the supported regions relative to the value in the unsupported gap region. For the upper sulfur layer we see a smaller and opposite variation with a potential valley in the gap region. This shows that charge transfer effects are dominant at this distance, in accordance with the results in Fig. 5. Figure 6 (d) shows the Mo $d_{z^2}$ orbital energies for both Au and Ti leads, which are the most important for the transport close to the CBM\cite{30}. While the Au substrate gives rise to a relatively smooth potential valley in the gap (due to the interaction with Au ions), the Mo $d_{z^2}$ orbital energies with the Ti substrate show larger variation and a potential hill in the gap region. The larger potential variations within the MoS$_2$ layer for a Ti substrate lead to stronger backscattering and thus explain the larger edge resistance for Ti electrodes.

In conclusion, we have theoretically investigated the electron transport through monolayer molybdenum dichalcogenide sheets suspended between metallic contacts. Two different kinds of contact resistances dominate the transport for different junction geometries. The interface resistance is large for small contact lengths and large metal-MoX$_2$ binding distances, while the edge resistance dominates for long contacts and small metal-MoX$_2$ binding distances. For sufficiently long contact length, which is likely the case in most experimental FET setups,\cite{23,41} electron transparency is enhanced with the weakly coupled metal/MoX$_2$ interface due to the minimal back-scattering arising from the smaller potential variation across the MoX$_2$ channel. In the particular case of MoS$_2$ with Au or Ti electrodes, we find that for short contacts (L$<$6 nm), Ti electrodes give the highest transmission while for longer contacts Au is the metal of choice. We expect these results and considerations to be general to many contacts between metals and 2D materials, and they should be taken into account when designing efficient metallic contact of 2D nanoelectronic devices.

\bibliography{main}
\end{document}